\documentclass[prd,aps,superscriptaddress,floatfix,nofootinbib,eqsecnum]{revtex4-2}

\pdfoutput=1

\usepackage{amsmath}
\usepackage{amsfonts}
\usepackage{amsmath}
\usepackage{amssymb}
\usepackage{bm}
\usepackage{dcolumn}
\usepackage{graphicx}
\usepackage[latin1]{inputenc}
\usepackage{latexsym}
\usepackage{rotating}
\usepackage{hyperref}
\usepackage{subfigure}
\usepackage{color}
\usepackage{changes}
\usepackage{verbatim}
\usepackage{comment}
\usepackage{empheq}
\usepackage{csquotes}
\usepackage{physics}
\usepackage{float}
\usepackage{soul}
\usepackage{amsmath,latexsym}
\usepackage{booktabs}  

\begin{document}

\title{Generalized Uncertainty Principle and the Zeeman Effect: Relativistic Corrections Unveiled}

\author{Gaurav Bhandari}\email{bhandarigaurav1408@gmail.com}\affiliation{Department of Physics, Lovely Professional University, Phagwara, Punjab, 144411, India}

\author{S. D. Pathak}\email{shankar.23439@lpu.co.in}\affiliation{Department of Physics, Lovely Professional University, Phagwara, Punjab, 144411, India}

\author{Manabendra Sharma}\email{sharma.man@mahidol.ac.th}\affiliation{Centre for Theoretical Physics and Natural Philosophy, Nakhonsawan Studiorum for Advanced Studies, Mahidol University, Nakhonsawan, 60130, Thailand}

\begin{abstract}
In this paper, we calculate the relativistic corrections to the Zeeman effect for hydrogen-like atoms based on the Generalized Uncertainty Principle (GUP). We propose a relativistic GUP algebra using the Stetsko and Tkachuk approximation and incorporate these corrections into the Zeeman effect. In the relativistic limit, our results recover previously derived GUP corrections as well as the standard Lande energy shift expression when GUP effects are absent. This work presents a generalized expression that accounts for both relativistic and GUP corrections to the Zeeman effect.
\end{abstract}

\maketitle

\textbf{Key words}: Generalized Uncertainty Principle, Zeeman effect, Relativistic corrections.



\section{Introduction}
The pursuit of a coherent and comprehensive theory of quantum gravity constitutes a paramount challenge in contemporary physics. Various frameworks, including String Theory (ST)\cite{veneziano1986stringy,witten1996reflections,scardigli1999generalized,gross1988string,amati1989can,yoneya1989interpretation,s3}, Loop Quantum Gravity (LQG) \cite{rovelli1998strings,sharma2019background,garay1995quantum,as1,ha1,sh1,Q2,Q3,Q4,Q5}, and Doubly Special Relativity (DSR) \cite{double1, gh1}, have emerged as potential candidates, each presenting distinct advantages. However, a comprehensive theory of quantum gravity remains elusive. A notable commonality among these approaches is the postulation of a minimal length scale. Specifically, in String Theory, this minimal length corresponds to the string length, whereas the integration of General Relativity (GR) with the Standard Model suggests a minimal length on the order of the Planck scale \cite{Z1,L0,L1,L2}. The proposed existence of a minimal length scale poses a contradiction to a fundamental tenet of quantum mechanics -the Heisenberg Uncertainty Principle (HUP). The HUP asserts that while the position of a particle can be measured with high precision, such precision necessitates a corresponding uncertainty in momentum. To reconcile this minimal length scale predicted by theories of quantum gravity \cite{Z2,Z3,Z4}, it becomes imperative to modify the HUP, thereby giving rise to the Generalized Uncertainty Principle (GUP).In the current landscape marked by the absence of a unified theory of quantum gravity, the Generalized Uncertainty Principle (GUP) has emerged as a significant framework. It encapsulates a shared characteristic inherent in numerous quantum gravity theories: the presence of a minimal length scale. This is accomplished by augmenting the Heisenberg uncertainty relation with an additional momentum-dependent term on the right-hand side, as expressed in the following formulation:
\begin{equation}
\Delta X \Delta P \geq \frac{\hbar}{2}\left[1+\beta_1 \left(\frac{l_p}{\hbar}\right)^2 (\Delta P)^2\right]
\end{equation}
where $l_p$ is the Planck length and $\beta_1$ is the positive numerical constant  \cite{e1,e2,bos1}. And, it is clear from the above relation that the minimal length $(\Delta X)_{min} = \sqrt{\beta_1} l_p $. 

The Generalized Uncertainty Principle (GUP) is grounded in the consideration of quantum fluctuations within space-time, which fundamentally restrict the precise, concurrent measurement of both position and momentum. These inherent uncertainties in the fabric of space-time introduce additional indeterminacy in the measurements of a particle's position and momentum, further reinforcing the limitations imposed by quantum mechanics \cite{tawfik2015review, lake2021generalised, PhysRevD.85.104029, ADLER_1999}.
Quantum fluctuations in space-time refer to the minute, spontaneous variations in the geometry of space-time at the quantum level. These fluctuations make it impossible to define precise positions and momenta simultaneously due to the inherent uncertainty in the space-time structure. The GUP formalizes this by extending the Heisenberg Uncertainty Principle to include the effects of gravitational and quantum forces, thereby leading to stricter limits on the precision of simultaneous measurements. The hypothesis that gravitational effects may alter the uncertainty principle was initially articulated by Mead in 1964 \cite{mead1964possible}. As the generalized uncertainty principle (GUP) modifies the Heisenberg uncertainty principle (HUP), it consequently influences fundamental aspects of quantum mechanics. Therefore, it is imperative to investigate the ramifications of GUP within established quantum mechanical frameworks to effectively evaluate theories of quantum gravity \cite{Z5,Z6,Z7,Z8}. Additionally, numerous phenomenological studies have been undertaken to assess the implications of GUP on cosmic dynamics \cite{la,lb,G1,G2,G3}. 

The predicted minimal length in quantum gravity is generally on the order of the Planck length ($l_{pl} = 10^{-35}$ m), where quantum gravitational effects are anticipated to become prominent. However, a significant challenge in integrating a minimal length into theoretical frameworks is its violation of Lorentz invariance, as the minimal length does not qualify as a Lorentz-invariant quantity. This constraint has limited the majority of generalized uncertainty principle (GUP) models to non-relativistic regimes, with only a few efforts made to extend GUP into relativistic contexts and Quantum Field Theory \cite{A1,A2,A3,A4,A5}. Recent research has investigated the generalized uncertainty principle (GUP) within a relativistic framework. Notably, \cite{B1} has introduced a relativistic generalized uncertainty principle (RGUP) that maintains Lorentz invariance while converging to the non-relativistic limit in low-energy scenarios. The establishment of this coherent relativistic GUP framework facilitates an examination of its implications for fundamental principles and elucidates how it modifies established results.

In this study, we explore the Zeeman effect as a quantum mechanical and atomic physics phenomenon within the framework of the RGUP. The Zeeman effect, which is thoroughly characterized in both non-relativistic and relativistic contexts, presents an ideal candidate for examining potential deviations arising from RGUP. High-precision measurements in atomic physics may yield indirect experimental evidence for generalized uncertainty principles or impose constraints on quantum gravity models, thereby providing a promising avenue for testing fundamental theoretical constructs.

The structure of this paper is organized as follows: In Sections (\ref{cla}) and (\ref{RGUP}), we present a comprehensive review of the non-relativistic Generalized Uncertainty Principle (GUP) and its corresponding algebraic framework. Additionally, we introduce the RGUP, as proposed by authors \cite{B1}, within the context of Minkowski spacetime, and we utilize the approximation method by Stetsko and Tkachuk \cite{B2} to derive modified canonical variables specifically, position and momentum that comply with the commutation relation $[x^{\mu}, x^{\nu}] = 0$. In Section (\ref{Zeeman}), we examine the implications of the RGUP on magnetostatic fields and the electromagnetic field tensor, further investigating its influence on the Zeeman effect. Finally, in Section \ref{con}, we provide a concise summary of our findings, delineate potential avenues for future research, and underscore the importance of analyzing the Zeeman effect in the framework of GUP.

\section{A brief review of Non-relativistic GUP}\label{cla}

In standard quantum mechanics, the position and momentum operators satisfy the commutation relation $ [x^i, p^j] = i \hbar \delta^{ij}$. To account for the presence of a minimal measurable length scale, Kempf and colleagues initially proposed a modified Heisenberg algebra \cite{Z2,Z3,Z4}. In the context of $D$-dimensional space, this modified algebra can be expressed as follows:

\begin{align}
[x^i, p^j] &= i \hbar \left[ (1 + a_1 p^2) \delta^{ij} + a_2 p^i p^j \right], \label{d1} \\
[x^i, x^j] &= i \hbar \frac{(2a_1 - a_2) + (2a_1 + a_2) a_1 p^2}{1 + a_1 p^2} (p^i x^j - p^j x^i), \\
[p^i, p^j] &= 0,
\end{align}
In this context, $a_1$ and  $a_2$  are non-negative deformation parameters that govern the algebraic structure within the deformed space, where $i, j = 0, 1, 2, \ldots, D$ . By applying Eq.~(\ref{d1}) in conjunction with the Schrodinger-Robertson uncertainty relation, one can ascertain the minimum uncertainty in position, expressed as follows:
\begin{equation}
(\Delta x^i)_{\text{min}} = \hbar \sqrt{(D a_1 + a_2)}.
\end{equation}

Using the commutator relation in Eq.~(\ref{d1}), Stetsko and Tkachuk presented approximate deformed variables to first order in the deformation parameters as:
\begin{align}
x^i &= x_0^i + \frac{2a_1 - a_2}{4} \left( p_0^2 x_0^i + x_0^i p_0^2 \right), \\
p^i &= p_0^i \left( 1 + \frac{a_2}{2} p_0^2 \right),
\end{align}
where $p^2 = p^i p^i$, and $x_0$ and $p_0$ are the position and momentum operators in the undeformed space (i.e., the standard position and momentum operators). In the special case where $a_2 = 2a_1$, such that $[x^i, x^j] = 0$, the commutation relations simplify to:
\begin{align}
[x^i, p^j] &= i \hbar \left[ (1 + a_1 p^2) \delta^{ij} + a_1 p^i p^j \right], \label{w1} \\
[x^i, x^j] &= 0, \label{w2} \\
[p^i, p^j] &= 0, \label{w3}
\end{align}
and the modified position and momentum variables that satisfy the relations (\ref{w1})-(\ref{w3}) to first order in $a_1$ are:
\begin{align}
x^i &= x_0^i, \\
p^i &= p_0^i \left( 1 + a_1 p_0^2 \right).
\end{align}

In the following section, we will apply a similar approximation to derive the algebra for the relativistic Generalized Uncertainty Principle (RGUP).

\section{Relativistic generalized uncertainty principle}\label{RGUP}
The quadratic form of GUP-modified commutator bracket between positon $x^i$ and momentum $p^j$ was first proposed by Kempf-Mangano-Mann \cite{Z3} as
\begin{equation}
[x^i,p^j]=i\hbar\delta^{ij}(1+ \beta_1 \textbf{p} ^2),\label{1}
\end{equation}
where $i,j \in 1,2,3$, and Eq.(\ref{1}) is non-relativistic. For this GUP the minimum bound in the uncertainty in position is calculated as $\Delta x_{min} = \hbar \sqrt{\beta_1}$. 
The most general covariant form of quadratic GUP in Minkowski space-time with signature $(- + + +)$, inspired by \cite{C1} is represented through the commutator bracket as
\begin{equation}
[x^{\mu}, p^{\nu}]=i \hbar(1+(\epsilon - \alpha)\gamma^2 p^{\rho}p_{\rho})\eta^{\mu \nu} +i \hbar (\beta+2\epsilon)\gamma^2 p ^{\mu}p^{\nu}, \label{2}
\end{equation}
where $\mu, \nu \in {0,1,2,3}$.In the above equation, we introduce the dimensional parameter $\gamma$, which has the dimension of inverse momentum, and the dimensionless parameters $\alpha$, $\epsilon$, $\beta$, and $\xi$ (the latter used below in Eq.~(\ref{eqx})). The parameter $\gamma$ is defined as $\gamma = \frac{1}{M_{Pl}c}$, where $M_{Pl}$ is the Planck mass. All of these parameters become relevant when quantum-gravitational effects are significant, particularly near the Planck scale. At the Planck scale, the dimensionless parameters are typically considered to be of order unity. For a more detailed study of these parameters, refer to \cite{ Z3,C1}.  

We note that Eq.(\ref{2}) reduces to Eq.(\ref{1}) for the non-relativistic ($c \rightarrow \infty$) limit and to standard Heisenberg Uncertainty relation as $\gamma \rightarrow 0$. Furthermore, $x^{\mu}$ and $p^{\nu}$ represent the physical position and momentum, which are not canonical conjugates. Therefore, we introduce new four-vectors, $x^{\mu}_0$ and $p^{\nu}_0$, such that
\begin{equation}
p^{\nu}_0 = -i \hbar \frac{\partial}{\partial x_{0 \nu}}, \quad  [x^{\mu}_0, p^{\nu}_0] = i \hbar \eta ^{\mu \nu}.
\end{equation}
Thus, the deformed position and momentum variables, expressed in terms of the new four vectors up to the order of $\gamma^2$, are written as
\begin{align}
  x^{\mu} &= x^{\mu}_0 - \alpha \gamma^2 p^{\rho}_0 p_{0 \rho}x^{\mu}_0 +\beta \gamma^2 p^{\mu}_0p^{\rho}_0 x_{0\rho}+ \xi \hbar \gamma^2p^{\mu}_0,\label{eqx}\\
    p^{\mu} &= p^{\mu}_0(1+\epsilon \gamma^2p^{\rho}_0p_{0\rho}). 
\end{align}
 The position operator obeys the following commutator relation as
 \begin{equation}
    [x^{\mu},x^{\nu}]= i \hbar \gamma^2 \frac{-2\alpha+\beta}{1+(\epsilon-\alpha)\gamma^2 p^{\rho}p_{\rho}}(x^{\mu}p^{\nu}-x^{\nu}p^{\mu}).
 \end{equation}
 In Eq.(\ref{eqx}) the last two terms break the isotropy of spacetime by by specifying a particular direction $p^{\mu}_0$.  As a result, this violates the principle of relativity, so we set $\beta$ and $\xi$ equal to 0. We can recover this relativistic GUP to the KMM algebra for $c \rightarrow \infty$. For the application of the Zeeman effect, we introduce the Stetsko and Tkachuk approximate representation to fulfill the KMM algebra in the first order by choosing $\alpha =0$.
 Now from the linear approximation, the modified algebra is 
 \begin{align}
 [x^{\mu}, p^{\nu}] &=i \hbar(1+\epsilon \gamma^2 p^{\rho}p_{\rho})\eta^{\mu \nu} + 2 i \hbar \epsilon \gamma^2 p ^{\mu}p^{\nu},\label{a1} \\
[x^{\mu}, x^{\nu}]&=0.\label{a2}
\end{align}
This holds true with the modified momentum and position as
\begin{align}
x^{\mu} &= x^{\mu}_0,\label{a3}\\
p^{\mu} &= p^{\mu}_0(1 + \epsilon \gamma^2 p^{\rho}_0p_{0 \rho}). \label{mp}
\end{align}
We can also derive the dispersion relation as the squared physical momentum $p^{\rho}p_{\rho}$ takes the following forms as
\begin{equation}
p^{\rho}p_{\rho}=-(mc)^2,
\end{equation}
or, in terms of $p^{\mu}_0$, we get
\begin{equation}
p^{\rho}_0p_{0\rho}(1+2 \epsilon p^{\sigma}_0p_{0\sigma}) = -(mc)^2,\label{3}
\end{equation}
where $m$ is the mass of the particle. Since the equation contains mixed derivatives (involving both space and time derivatives, as well as combinations of the two), it becomes more complicated to solve analytically, even in the spherically symmetric case. Therefore, we solve for $p^{\rho}_0 p_{0\rho}$ from Eq.~(\ref{3}), focusing on the solutions that reduce to $(mc)^2$ in the $\gamma \rightarrow 0$ limit. This gives us the effective second-order dispersion relation from the quadratic formulae as
\begin{align}
p^\rho_0 p_{0\rho} &= -\frac{1}{4 \epsilon \gamma^2} + \sqrt{\frac{1}{(4 \epsilon \gamma^2)^2} - \frac{(mc)^2}{2 \epsilon \gamma^2}}, \nonumber \\
&\simeq - (mc)^2 - 2 \epsilon \gamma^2 (mc)^4 - \mathcal{O}(\gamma^4). 
\label{rmo}\end{align}
We can also recover the non-relativistic four-momentum scalar product having a relation 
$p^\rho_0 p_{0\rho} = -\left(\frac{E}{c}\right)^2 + p^i_0 p_{i_0}$  by taking the limit $c \rightarrow \infty$ as
\begin{equation}
p^\rho_0 p_{0\rho} = - \hbar^2 \grad^2. \label{b2}
\end{equation}
In the next section, we apply the above results of RGUP to get the modified Zeeman effect and modified energy shift for hydrogen or hydrogen-like atoms in a uniform magnetostatic field.

\section{Zeeman Effect in presence of RGUP}\label {Zeeman}

To study the effect of relativistic correction of GUP in the Zeeman effect, we first see the effect in the magnetic field. To start with, the relativistic Lagrangian field density  with current density is given 
\begin{equation}
\mathcal{L}= - \frac{1}{4 \mu _0}F_{\mu \nu}F^{\mu \nu} - A_{\mu} \mathcal{J}^{\mu}
\end{equation}
where $\mu, \nu = 0,1,2,3$ and $F_{\mu \nu}(x)=\partial_{\mu}A_{\nu}(x)-\partial_{\nu}A_{\mu}(x)$ and $\textbf{A}(x)=(A^0(x),A^1(x),A^2(x),A^3(x))$ are electromagnetic field tensor and vector potential respectively. Now using the deformed position and derivative operator in Eq.(\ref{mp}) as
\begin{equation}
x^\mu\rightarrow X^\mu \equiv x^\mu_0,\quad
\partial^{\mu} \rightarrow D^{\mu}:= \partial^{\mu}_0(1-\epsilon \gamma^2(mc)^2-\mathcal{O}(\gamma^4)), \label{eq}
\end{equation}
where $D^{\mu}$ is the ordinary derivative for Minkoskian space-time. From this point onward, we will use \textbf{boldface} letters to denote four-vectors, which include both spatial and temporal components. Using Eq.(\ref{eq}) the modified electromagnetic field tensor become
\begin{align}
\mathcal{F}_{\mu \nu}(\mathbf{X})&=D_{\mu} A_{\nu}(\textbf{X}) - D_{\nu}A_{\mu}(\textbf{X}), \nonumber \\
\mathcal{F}_{\mu \nu}(\textbf{x})&= F_{\mu \nu}(\textbf{x}) - \epsilon \gamma^2 (mc)^2F_{\mu \nu}(\textbf{x}) - \mathcal{O}(\gamma^4).
\label{f}\end{align}
We define the magnetic induction vector  and electric field induction $\textbf{B}(\textbf{x})$, $\textbf{E}(\textbf{x})$ as
\begin{equation}
F_{ij}=-\epsilon_{ijk}B^k, \quad F^{ij}=\epsilon^{ijk}B_k, \quad F_{i0}=E^i\label{g}
\end{equation}
since we are studying the effect of magnetic field in the atomic spectra we consider the electric field vector to be zero. The magnetic induction vector in three-dimensional space with $i, j = 1,2,3$ is 
\begin{equation}
B^i=\{B_x,B_y,B_z\}.\label{b}
\end{equation}
Using Eq.(\ref{g}), Eq.(\ref{b}) and Eq.(\ref{f}), one can obtain the GUP-modified magnetostatic field as
\begin{align}
\textbf{b}_{RGUP}(\textbf{x})
&= (1- \epsilon \gamma^2 (mc)^2 - \mathcal{O}(\gamma^4))\textbf{B}(\textbf{x}), \label{bgup}
\end{align}
 The above expression includes the effect of relativistic GUP and as we choose $c\rightarrow \infty $ the effects became non-relativistic and the expression up to the order of $\gamma^2$ became 
 \begin{align}
\textbf{b}_{GUP}(\textbf{x}) &= (1-\epsilon \gamma^2\hbar^2 \nabla ^2)\textbf{B}(\textbf{x})\\
&= (1 +\epsilon \gamma^2 p^2)\textbf{B}(\textbf{x})
 \end{align}
and as quantum gravity parameter $\gamma \rightarrow 0$ we recover our original magnetic field. We also note that in the relativistic limit ($c\rightarrow \infty$) we recover the results presented in \cite{C2}. 
\subsection{Zeeman effect with relativistic GUP}
The Hamiltonian $H_0$ for hydrogen-like atoms on taking the relativistic correction only in the kinetic energy term is 
\begin{equation}
H_0= \frac{p^2}{2m_e}-\frac{p^4}{8m_ec^2} + U^{(0)},
\end{equation}
where $U^{(0)}$  are the internal interactions of the zeroth order in $v/c$. We derive the magnetostatic field $\textbf{B}$ from the vector potential as
\begin{equation}
\textbf{A}= \frac{1}{2}(\textbf{B} \cross \textbf{X}).
\end{equation}
To incorporate the effect of an external magnetic field on the charged particle, we use the minimal coupling prescription. In the presence of an electromagnetic field, the canonical momentum $\textbf{p}$ is replaced by
\begin{equation}
\textbf{p}\rightarrow \textbf{p}- \frac{e\textbf{A}}{c}. \label{p}
\end{equation}
In the process of obtaining the Hamiltonian in the presence of relativistic GUP, we first find the modified vector potential. From Eq.(\ref{bgup}) we get
\begin{equation}
\textbf{A}_{RGUP} = \frac{1}{2}(\textbf{b}_{RGUP} \cross \textbf{x})= \frac{1}{2}(1- \epsilon \gamma^2 (mc)^2)(\textbf{B} \cross \textbf{x}),
\end{equation}
where \textbf{B} represents the usual magnetostatic field without GUP. On choosing \textbf{B} in positive z-direction, we have
\begin{align}
\textbf{A}_{RGUP}&=-\frac{1}{2}(1- \epsilon \gamma^2 (mc)^2)(By \hat{x}-Bx \hat{y}) ,\nonumber \\
&= (1- \epsilon \gamma^2 (mc)^2) \textbf{A}, \label{AR}
\end{align}
where $\textbf{A}=-\frac{1}{2}(By \hat{x}-Bx \hat{y})$. And, Eq.(\ref{p}) modifies using Eq.(\ref{AR}) and Eq.(\ref{mp}) as
\begin{equation}
\textbf{p}\rightarrow\textbf{P}- \frac{e \textbf{A}_{RGUP}}{c} = (1- \epsilon \gamma^2 (mc)^2)\left(\textbf{p}- \frac{e\textbf{A}}{c}\right). 
\end{equation}
Now, the relativistic Hamiltonian is 
\begin{equation}
H_{RGUP}= \frac{1}{2m_e}\left\{(1- \epsilon \gamma^2 (mc)^2)\left(\textbf{p}-\frac{e\textbf{A}}{c}\right)\right\}^2 - \frac{1}{8m_e^3c^2}\left\{(1- \epsilon \gamma^2 (mc)^2)\left(\textbf{p}-\frac{e\textbf{A}}{c}\right)\right\}^4
+ U^{(0)}\end{equation}
after neglecting order of $\gamma^4$, we obtain
\begin{equation}
\begin{split}
H_{RGUP} &= \frac{(1-2\epsilon \gamma^2 (mc)^2)}{2m_e} \left\{ \textbf{p}^2 + \frac{e^2\textbf{A}^2}{c^2} 
- \frac{e}{c} (\textbf{p} \cdot \textbf{A} + \textbf{A} \cdot \textbf{p}) \right\} \\
&\quad - \frac{(1+4 \epsilon \gamma^2 (mc)^2)}{8m_e^3c^2} \left\{ \textbf{p}^4 + \frac{e^4 \textbf{A}^4}{c^4} 
+ \frac{2\textbf{p}^2 e^2 \textbf{A}^2}{c^2} + \frac{e^2}{c^2} (\textbf{p} \cdot \textbf{A} + \textbf{A} \cdot \textbf{p})^2 \right. \\
&\quad \left. - \left(\frac{2e\textbf{p}^2 }{c} + \frac{2e^3 \textbf{A}^2}{c^3} \right) 
(\textbf{p} \cdot \textbf{A} + \textbf{A} \cdot \textbf{p}) \right\} + U^{(0)}.
\end{split}
\label{fham}
\end{equation}
We know that in Coulomb gauge $(\grad . \textbf{A}=0)$, we can replace \textbf{p}.\textbf{A} by \textbf{A}. \textbf{p}. We also have 
\begin{align}
\textbf{A}.\textbf{p}&=\frac{B}{2}\left(-yp_x+xp_y\right)=\frac{1}{2}BL_z,\\
\textbf{A}^2&=\frac{1}{4}B^2(x^2+y^2).
\end{align}
Since the relativistic GUP-modified Hamiltonian contains terms with \( \mathbf{A}^2 \), these can be neglected because of the quadratic term \( B^2(x^2 + y^2) \) that corresponds to higher-order magnetic corrections. In a one-electron system like hydrogen, these quadratic terms are much smaller than the linear terms, as the magnitude of the magnetic interaction for a single electron is relatively weak, and the quadratic terms contribute a much smaller energy shift.The Hamiltonian consists of both an unperturbed and a perturbed part. The perturbation arises from two contributions: \( H_B \), due to the external magnetic field, and \( H_{LS} \), due to the spin-orbit (\(\mathbf{L} \cdot \mathbf{S}\)) coupling. In the first case, we consider only the contribution from a strong magnetic field, where the external magnetic field is strong enough to dominate over the internal spin-orbit coupling. As a result, the interaction between the spin magnetic moment and the external magnetic field becomes significant. The final effective Hamiltonian in the first case will take the form as
\begin{equation}
H_{RGUP}= \frac{p^2}{2m_e}-\frac{\epsilon \gamma^2 (m c)^2 p^2}{m_e}-(1-2\epsilon \gamma^2 (mc)^2) \frac{e}{2m_ec}BL_z- \frac{(1+4 \epsilon \gamma^2 (mc)^2)}{8m_e^3c^2} p^4 +\frac{e}{4m_e^3 c^3}BL_zp^2 +U^{(0)}.
\end{equation}
We also add the spin magnetic-moment interaction in the presence of a relativistic effect \cite{j,i} with RGUP as
\begin{equation}
-\mu. \textbf{b}_{RGUP}= -\frac{e}{m_e c}\left(1+\frac{\alpha'}{2\pi}\right)\textbf{s}.(1- \epsilon \gamma^2 (mc)^2 - \mathcal{O}(\gamma^4))\textbf{B} \simeq  -\frac{e}{m_e c}\left(1+\frac{\alpha'}{2\pi}\right)(1- \epsilon \gamma^2 (mc)^2) Bs_z,
\end{equation}
where $\alpha '$ is the fine-structure constant and this correction accounts for the relativistic. This gives the total modified  Hamiltonian as
\begin{align}
H_{0} &= \frac{p^2}{2m_e} - \frac{p^4}{8m_e^3c^2} + U^{(0)} \nonumber \\
H^{\text{RGUP}}_B &= -\frac{\epsilon \gamma^2 (mc)^2 p^2}{m_e} 
- \left(1 - \epsilon \gamma^2 (mc)^2 \right) \frac{eB}{2m_ec}(L_z + 2s_z) 
+ \frac{eB}{2m_ec} \epsilon \gamma^2 (mc)^2 L_z \nonumber \\
&\quad + \frac{eB}{4m_e^3 c^3} L_z p^2 
- \frac{\alpha'}{2\pi} \left(1 - \epsilon \gamma^2 (mc)^2 \right) \frac{e}{m_e c} B s_z 
- \frac{\epsilon \gamma^2(mc)^2}{2m_e^3c^2}p^4.
\end{align}

The Relativistic GUP correction to the Hamiltonian acts as a small perturbation. To determine the effects of the RGUP corrections, we use the $J^2$, $J_z$ eigenkets as our basis kets. We will assume that the eigenvalues are not affected by the GUP corrections, allowing us to study the leading-order corrections. The energy shift in the first-order in the deformed space is 
\begin{equation}
\begin{aligned}
&-\frac{eB}{2m_e c} \langle J_z + S_z \rangle_{j=L \pm \frac{1}{2}, m'} 
- \frac{\alpha' eB}{2 \pi m_e c} \langle S_z \rangle_{j=L \pm \frac{1}{2}, m'} + \frac{eB}{4 m_e^3 c^3} \langle p^2 (J_z - S_z) \rangle_{j=L \pm \frac{1}{2}, m'} \\
&+ \epsilon \gamma^2 (mc)^2 \left[ \frac{eB}{2 m_e c} \langle J_z + S_z \rangle_{j=L \pm \frac{1}{2}, m'} 
+ \frac{eB}{2 m_e c} \langle J_z - S_z \rangle_{j=L \pm \frac{1}{2}, m'} 
+ \frac{eB \alpha'}{2 m_e c \pi} \langle S_z \rangle_{j=L \pm \frac{1}{2}, m'} 
- \frac{\langle p^2 \rangle}{m_e} + \frac{\langle p^4\rangle}{2m_e^3c^2}\right],
\end{aligned}\label{ra}
\end{equation}
in Eq.(\ref{ra}), we note that if we disregard relativistic effects, the second and third terms will vanish, and the factor $(mc)^2$ will change to $-p^2$, retaining the expression for the standard GUP correction as discussed in \cite{C2}.
The expectation value of \(J_z\) is \(m' \hbar\) where  $m'$ is the magnetic quantum number, and from the standard quantum mechanical results, we get
\begin{equation}
\langle S_z \rangle_{j=L \pm \frac{1}{2}, m'} = \pm \frac{m' \hbar}{2l+1}.
\end{equation}
We also know that for \(\langle p^2 \rangle\) and \(\langle p^4 \rangle\), we use
\begin{equation}
    \nabla^2 = \left(\frac{\partial^2}{\partial r^2} + \frac{2}{r} \frac{\partial}{\partial r} - \frac{\textbf{L}^2}{\hbar^2 r^2}\right).
\end{equation}
Using the above expressions we obtain Lande's formulae for energy shift $\Delta \mathcal{E}_B$ in the presence of RGUP up to the order of $\mathcal{O}(\gamma^2)$ is obtained as

\begin{align}
(\Delta \mathcal{E}_B)_{\text{RGUP}} = & -\frac{eB\hbar}{2m_e c} m' \left[1 \pm \frac{1}{2l+1}\right] 
\mp \frac{\alpha' eB}{2 \pi m_e c} \frac{m' \hbar}{2l+1} 
+ \frac{eB}{4 m_e^3 c^3} \frac{m' \hbar^2}{r_0^2} l(l+1) \left(1 \mp \frac{1}{2l+1}\right) \nonumber \\
& + \epsilon \gamma^2 (mc)^2 \Bigg[ \frac{eB}{2m_e c} m' \hbar \left(1 \pm \frac{1}{2l+1}\right)  
+ \frac{eB}{2m_e c} m' \hbar \left(1 \mp \frac{1}{2l+1}\right) \mp \frac{eB \alpha'}{2 m_e c \pi} \frac{m' \hbar}{2l+1}  \nonumber \\
& 
- \frac{\hbar^2 l(l+1)}{m_e} 
+ \frac{\hbar^4}{2m_e^3c^2 r_0^4} (l(l+1))^2 \Bigg]. \label{finaleq}
\end{align}

where \(r_0\) is the Bohr radius. The second line in Eq.~(\ref{finaleq}) shows the correction to the energy shift due to the RGUP. As the effects of relativity and GUP vanish, we recover the original result for the energy shift, which is proportional to \(m'\) in \cite{C2} as
\begin{equation}
(\Delta \mathcal{E}_B)_{\text{RGUP}} =  -\frac{eB\hbar}{2m_e c} m' \left[1 \pm \frac{1}{2l+1}\right] 
 + \frac{\epsilon \gamma^2}{m_e} \Bigg[ \frac{\hbar^4}{ r_0^4} (l(l+1))^2 - \frac{eBm'}{ c}  \frac{\hbar^2}{ r_0^2} l(l+1)\left(1 \pm \frac{1}{2l+1}\right)  
 \Bigg]. \label{GH}
\end{equation}
With the presence of RGUP, the energy shift is no longer proportional to the magnetic quantum number \(m'\) due to the inclusion of the last terms \(\langle p^2 \rangle\) and \(\langle p^4 \rangle\)  in Eq.~(\ref{finaleq}).

Now, if we consider the effect of \(\mathbf{L} \cdot \mathbf{S}\) coupling, we first derive the RGUP-modified Hamiltonian associated with it. For a central (spin-independent) potential \(U^{(0)}\), the electron's magnetic moment interacts with the effective magnetic field generated by the moving charge in the electric field. Additionally, accounting for the extra contribution to the energy from Thomas precession, in conjunction with relativity, reduces the factor of spin-orbit coupling by \(1/2\) (known as the Thomas factor) \cite{C3,D1,Q1}, yielding a modified interaction.

\begin{equation}
H^{\text{RGUP}}_{LS} =\left(\frac{e \textbf{S}}{m_e c}\right) \cdot \left[\frac{1}{2 m_ec}\frac{1}{r} (\textbf{P} \cross \textbf{x}) \frac{d U^{(0)}}{dr}\right],
\end{equation}
and using Eq.(\ref{mp}) we get
\begin{align}
 H^{\text{RGUP}}_{LS} &= \left(\frac{e \textbf{S}}{m_e c}\right) \cdot \left[\frac{(1-\epsilon \gamma^2 (mc)^2)}{2 m_ec}\frac{ (\textbf{p} \cross \textbf{x})}{r} \frac{d U^{(0)}}{dr} \right] \nonumber \\
 &= \frac{(1-\epsilon \gamma^2 (mc)^2 )}{2 m_e^2c^2 } \frac{1}{r} \frac{d U^{(0)}}{dr} (\textbf{L} \cdot \textbf{S}). 
\end{align}
For the above expression we use the $L_z$ and $S_z$ eigenkets $|l,s = \frac{1}{2}, m_l, m_s\rangle$ are the base kets. The expectation value of $\textbf{L} \cdot \textbf{S}$ with respect to $| m_l, m_s \rangle$ is 
\begin{align}
\langle \textbf{L} \cdot \textbf{S} \rangle  &= \langle L_z S_z + \frac{1}{2}(L_+ S_- + L_- S_+) \rangle _{m_l m_s} \nonumber \\
&= \hbar^2 m_l m_s,
\end{align}
where we used 
\begin{equation}
\langle L_{\pm} \rangle _{m_l} = 0 , \quad \langle S_{\pm} \rangle_{m_s} = 0,
\end{equation}
so the expectation value of the Hamiltonian approximate to the leading-order correction became
\begin{equation}
\langle H^{\text{RGUP}}_{LS} \rangle =  \frac{(1-\epsilon \gamma^2 (mc)^2 ) \hbar^2 m_l m_s}{2 m_e^2c^2 } \left\langle \frac{1}{r} \frac{d U^{(0)}}{dr} \right\rangle . \label{HLS}
\end{equation}
For non-relativistic ($c \rightarrow \infty$) and without GUP ($\gamma \rightarrow 0$), Eq.(\ref{HLS}) became the usual $\langle H_{LS} \rangle$ with value equal to $ \frac{ \hbar^2 m_l m_s}{2 m_e^2c^2 } \left\langle \frac{1}{r} \frac{d U^{(0)}}{dr} \right\rangle$. In this case, the perturbed Hamiltonian contains two terms one from the external magnetic field and one from the \textbf{L}.\textbf{S} coupling.We observe that the presence of RGUP does not alter the number of Zeeman lines, but it modifies the energy shift spectrum. Furthermore, when comparing the energy shift relations for GUP and RGUP, as shown in Eq. (\ref{GH}) and Eq. (\ref{RGUP}), respectively, we find that the modified terms in the GUP case are proportional to the orbital angular momentum quantum number \( l \). However, in the case of RGUP, they depend on a combination of \( m' \) and \( l \). This implies that for the \( s \)-orbital singlet state (\( l = 0 \)), where \( m' \) is also zero, there is no correction to the energy shift for either GUP or RGUP. For orbitals with \( l \neq 0 \), both GUP and RGUP introduce additional energy shifts. Specifically, the shifts depend on $l$, $m'$, and higher-order relativistic corrections, leading to distinct shifts for different levels. These corrections are particularly visible in higher angular momentum states, where they contribute significantly to Lande's energy shift.

\section{Conclusion} \label{con}

The study of relativistic generalized uncertainty principles (GUP) is essential for advancing our understanding of quantum gravity, particularly at Planck-scale energies (approximately \(10^{22}\) MeV) where its effects become increasingly pronounced. The interplay of relativistic and quantum mechanical phenomena at high energy scales underscores the necessity for a comprehensive framework that integrates these aspects. Relativistic GUP serves as this framework, enabling the retrieval of GUP-modified effects when relativistic corrections are factored in. These GUP and RGUP corrections hold particular significance for applications in astrophysics and high-energy physics. In extreme environments, such as the intense magnetic fields surrounding neutron stars and magnetars \cite{n1,n2,n3}, both relativistic and quantum gravity effects are crucial. Studying the atomic spectra in proximity to these celestial objects can yield valuable insights into the influence of quantum gravity on spectral lines, providing a promising avenue for testing various theories of quantum gravity.

In this paper, we explore the impact of the Relativistic Generalized Uncertainty Principle (RGUP) on the Zeeman effect in hydrogen-like atoms, specifically by applying relativistic corrections solely to the kinetic term of the Hamiltonian. 
We introduced a relativistic GUP algebra, leveraging the Stetsko and Tkachuk approximation\cite{B2}, which adheres to the modified algebra given by Eqs.~(\ref{a1})-(\ref{mp}).  Our findings culminated in a valid four-momentum solution in the relativistic framework, as presented in Eq. (\ref{rmo}). Notably, this solution seamlessly transitions to the non-relativistic squared momentum in the limit as \(c \to \infty\), as shown in Eq. (\ref{b2}). This work not only advances our understanding of the RGUP but also provides significant insights into its implications for the Zeeman effect, paving the way for further investigations in quantum mechanics. Our findings reveal the RGUP-modified Lande energy shift, which diverges from the standard theory by not exhibiting a direct proportionality to the magnetic quantum number $m'$. Upon eliminating the relativistic correction, we successfully recover the GUP-modified Zeeman effect, as previously shown in \cite{C2}. Moreover, by disregarding the GUP correction parameter $\gamma $, we revert to the standard Zeeman effect. The introduction of RGUP introduces a more intricate energy shift, resulting in an expanded range of energy shifts, which can lead to variations that slightly elevate or diminish the overall energy shift. This complexity highlights the nuanced interplay between RGUP effects and magnetic interactions, paving the way for further exploration in this area of research. Although our study focuses specifically on the Zeeman effect, the methodologies employed can be readily adapted to investigate other atomic phenomena, including the Stark effect. Additionally, our analysis has been conducted within the framework of a flat Minkowski spacetime. To achieve a more comprehensive and physically accurate understanding, it is imperative to extend this research into curved spacetime, where the presence of non-zero affine connections must be taken into account. We propose this exploration as a promising avenue for future research.



\end{document}